\documentclass{JHEP3}%
\usepackage{amsfonts}
\usepackage{amsmath}
\usepackage{amssymb}
\usepackage{graphicx}
\usepackage{epsfig}%
\newtheorem{theorem}{Theorem}

\title{The number of eigenstates: counting function and heat kernel}
\author{Wu-Sheng Dai and Mi Xie\\Department of Physics, Tianjin University, Tianjin 300072, P. R. China\\LiuHui Center for Applied Mathematics, Nankai University \& Tianjin
University, Tianjin 300072, P. R. China\\
E-mail: \email{daiwusheng@tju.edu.cn}\\
E-mail: \email{xiemi@tju.edu.cn}}

\abstract{The main aim of this paper is twofold: (1) revealing a
relation between the counting function $N\left(  \lambda\right)  $
(the number of the eigenstates with eigenvalue smaller than a given
number) and the heat kernel $K\left( t\right)  $, which is still an
open problem in mathematics, and (2) introducing an approach for the
calculation of $N\left(  \lambda\right)  $, for there is no
effective method for calculating $N\left(  \lambda\right)  $ beyond
leading order. We suggest a new expression of $N\left(
\lambda\right) $ which is more suitable for practical calculations.
A renormalization procedure is constructed for removing the
divergences which appear when obtaining $N\left(  \lambda\right)  $
from a nonuniformly convergent expansion of $K\left(  t\right)  $.
We calculate $N\left(  \lambda\right)  $ for $D$-dimensional boxes,
three-dimensional balls, and two-dimensional multiply-connected
irregular regions. By the Gauss-Bonnet theorem, we generalize the
simply-connected heat kernel to the multiply-connected case; this
result proves Kac's conjecture on the two-dimensional
multiply-connected heat kernel. The approaches for calculating
eigenvalue spectra and state densities from $N\left(  \lambda\right)
$ are introduced.}
\keywords{Differential and Algebraic Geometry, Boundary Quantum
Field Theory}
\preprint{}
\begin{document}

\section{Introduction}

A problem stemming from physics and soon becoming an important mathematical
problem is to recover geometry of a manifold from the knowledge of the
eigenvalues of a natural differential operator. This problem originates in the
theory of radiation (how to determine the state density of the electromagnetic
wave in a given cavity) and is formulated by Kac as "Can one hear the shape of
a drum?" \cite{Kac}.

What the original problem asked is that how many eigenvalues are smaller than
a given number $\lambda$. This problem is formulated as to seek the so-called
counting function $N\left(  \lambda\right)  $. For a given spectrum $\left\{
\lambda_{n}\right\}  $, the counting function is directly defined to be%
\begin{equation}
N\left(  \lambda\right)  =\text{the\ number of eigenvalues (with multiplicity)
smaller than }\lambda\text{,}\label{NLmd}%
\end{equation}
i.e.,%
\begin{equation}
N\left(  \lambda\right)  =\sum_{\lambda_{n}<\lambda}1.\label{NLsgm1}%
\end{equation}
In other words, the counting function represents the number of the eigenstates
whose eigenvalues are smaller than $\lambda$. Nevertheless, the counting
function $N\left(  \lambda\right)  $ is very difficult to calculate and there
is no general method for calculating $N\left(  \lambda\right)  $ in
mathematics \cite{Berger}. This is because when calculating $N\left(
\lambda\right)  $, one often encounters some unsolved problems in number
theory. For example, when calculating the counting function for the spectrum
of the Laplace operator on a tori, one encounters the Gauss circle problem in
number theory. The Hardy-Littlewood-Karamata Tauberian theorem gives the first
term of the asymptotic expansion of $N\left(  \lambda\right)  $, but does not
provide any information beyond the first-order term \cite{Berger}.

For a given spectrum $\left\{  \lambda_{n}\right\}  $, one also introduces the
heat kernel,
\begin{equation}
K\left(  t\right)  =\sum_{n}e^{-\lambda_{n}t},\label{Kt}%
\end{equation}
which is an another important function describing the relation between the
eigenvalues of the operator and the geometrical property of the manifold. The
heat kernel is relatively easy to calculate and some methods for calculating
heat kernels have been developed \cite{Vas}. In special, Elizalde \textit{et
al.} developed a very effective approach for calculating $K\left(  t\right)  $
\cite{ER,Eli,BEK,BEKL,EBK}.

There is no doubt that there must exist a relation between the counting
function $N\left(  \lambda\right)  $ and the heat kernel $K\left(  t\right)
$, since they are both determined by the same spectrum $\left\{  \lambda
_{n}\right\}  $. Nevertheless, the relation between $N\left(  \lambda\right)
$ and $K\left(  t\right)  $ is still an open problem in mathematics
\cite{Berger}.

Though the definition of counting function $N\left(  \lambda\right)  $, eq.
(\ref{NLmd}) or (\ref{NLsgm1}), is comprehensive and intuitive, other than the
definition of heat kernel $K\left(  t\right)  $, eq. (\ref{Kt}), the
definition of $N\left(  \lambda\right)  $ is not suitable for practical
calculations. In fact, if starting from this definition to calculate the
counting function, one may always encounter the mathematical difficulty
mentioned above. In this paper, we will propose an expression for $N\left(
\lambda\right)  $, which has a similar form with the definition formula of
$K\left(  t\right)  $, eq. (\ref{Kt}), and is more operable than the
definition (\ref{NLsgm1}). This expression can be regarded as an alternative
definition formula for $N\left(  \lambda\right)  $.

As a main result of this paper, we provide a relation\ between the counting
function $N\left(  \lambda\right)  $ and the heat kernel $K\left(  t\right)
$. This problem interests us from both purely mathematical viewpoint and
practical viewpoint. This relation allows us to calculate $N\left(
\lambda\right)  $ from $K\left(  t\right)  $ which, as mentioned above, is
often relatively easier to calculate.

The relation between the counting function $N\left(  \lambda\right)  $ and the
heat kernel $K\left(  t\right)  $ presented in this paper is an integral
transformation. However, in practice the heat kernel is often given in the
form of a series expansion and in most cases only the first several expansion
coefficients can be obtained \cite{Vas}. This requires us to integrate term by
term. Nevertheless, the series expansion of $K\left(  t\right)  $ is not
convergent uniformly. As a result, the integral of some terms will be
divergent. The counting function $N\left(  \lambda\right)  $ is of course
finite; the divergences are caused by illegally integrating a nonuniformly
convergent series term by term. For dealing with this problem, we provide a
renormalization procedure to remove the divergences.

In fact, the results of this paper provide two approaches for calculating the
counting function $N\left(  \lambda\right)  $: (1) by the expression for
$N\left(  \lambda\right)  $ presented in this paper, one can calculate
$N\left(  \lambda\right)  $ directly, and (2) by the relation between
$N\left(  \lambda\right)  $ and $K\left(  t\right)  $, one can first calculate
the heat kernel $K\left(  t\right)  $ and then calculate $N\left(
\lambda\right)  $ from $K\left(  t\right)  $. As a comparison, we calculate
$N\left(  \lambda\right)  $ for a $D$-dimensional box by both these two
approaches, respectively.

As an example of the calculation of counting functions by the relation between
$N\left(  \lambda\right)  $ and $K\left(  t\right)  $, we calculate $N\left(
\lambda\right)  $ from $K\left(  t\right)  $ for three-dimensional balls.

As another example, we calculate the counting function from the heat kernel
for the minus Laplace operator in a two-dimensional region with irregular
shape and nontrivial topology. For two-dimensional heat kernels, there are two
known results: the heat kernel for a simply-connected region \cite{NPD} and a
hypothesis made by Kac on the heat kernel for the multiply-connected case
\cite{Kac}. In the present paper, we first generalize the heat kernel for a
simply-connected region given in \cite{NPD} to the case of a
multiply-connected region which is bounded by a smooth but irregular curve and
with some holes. This result proves Kac's conjecture. Then, we calculate the
counting function for the two-dimensional multiply-connected region from the
heat kernel by the relation between $N\left(  \lambda\right)  $ and $K\left(
t\right)  $.

As applications, we discuss the problem that how to calculate the asymptotic
expressions for eigenvalue spectra and state densities from the counting
function $N\left(  \lambda\right)  $. By $N\left(  \lambda\right)  $, we
construct an equation for eigenvalues and then provide an approximate solution
in which the eigenvalue is expressed as a function of heat kernel
coefficients. Moreover, we also provide two expressions for state densities by
the relation between $N\left(  \lambda\right)  $ and $K\left(  t\right)  $ and
by the expression of $N\left(  \lambda\right)  $ provided in the present paper.

The counting function and the heat kernel are interesting in both mathematics
and physics:

In mathematics, the relation between the spectrum of the Laplace operator on a
Riemannian manifold and the geometry of this Riemannian manifold is an
important subject \cite{Berger,Berard,Milnor,GWW}, and the problem of spectral
asymptotics is one of the central problems in the theory of partial
differential operators \cite{Ivrii}. One of the main problems is to seek the
asymptotic expansions of the counting function $N\left(  \lambda\right)  $ and
the heat kernel $K\left(  t\right)  $. The general relation between $N\left(
\lambda\right)  $ and $K\left(  t\right)  $ is still unknown. Especially,
there is no general method for calculating the asymptotic expansion of the
counting function $N\left(  \lambda\right)  $.

In physics, the spectrum of the Laplace operator on a Riemannian manifold can
be directly applied to boundary quantum fields. Moreover, reconstructing the
geometrical property of a system from an eigenproblem is an interesting and
important problem. For example, Aurich \textit{et al.} reconstruct the shape
of the universe from the result of the eigenproblem \cite{ALSH}. The counting
function is directly related to the Casimir effect, the spectrum problem, and
the state density, etc. Moreover, there are many studies on heat kernels. In
quantum field theory, it is of crucial importance to evaluate the one-loop
divergences, and the control of the ultra-violet divergences can be achieved
by using heat kernel regularization methods \cite{Vas,BCP,Avr}. The heat
kernel expansion becomes a standard tool in the calculations of vacuum
energies \cite{BK}, the Casimir effect \cite{Mar,EMN}, and quantum anomalies
\cite{MV}. Now a large amount of research has been devoted to quantum gravity
based on the heat kernel expansion, including semiclassical approaches
\cite{Sha}, black hole thermodynamics \cite{CO}, wormhole physics \cite{KB},
heat kernels in curved space \cite{Sal}, and supergravity theories
\cite{BH,Hoo}. In addition, the heat kernel expansion is also used to study
string theory \cite{KKS,BN,MSN} and noncommutative field theory
\cite{FGV,ABBP,GIV}. The relation between the counting function and the heat
kernel allows one to introduce the concept of the counting function into these
fields. Moreover, there exist some indirect ways to construct the heat kernel
without solving the heat equation, and the heat kernel coefficients for
various manifolds have been discussed \cite{Vas,BEK,PN,DK}, which makes the
heat kernel expansion a very convenient and effective tool. By the relation
provided in the present paper, such results of the heat kernel can be directly
applied to the problem of the counting function.

In section \ref{definitionofNL}, we provide an expression for the counting
function $N\left(  \lambda\right)  $, which is more operable compared to the
definition (\ref{NLsgm1}) and can be regarded as an alternative definition
formula for the counting function. In section \ref{relation}, we point out a
relation between the counting function $N\left(  \lambda\right)  $ and the
heat kernel $K\left(  t\right)  $. In section \ref{calculating}, we provide a
renormalization procedure for removing the divergences appearing in the
calculation of $N\left(  \lambda\right)  $ from a nonuniformly convergent
series expansion of $K\left(  t\right)  $. In section \ref{D-box}, we compare
the two approaches for calculating $N\left(  \lambda\right)  $ using a
$D$-dimensional box as an example. In section \ref{3Dballs}, we calculate the
counting function of a three-dimensional ball from the heat kernel. In section
\ref{2Dmul}, for two-dimensional cases, we first generalize the heat kernel in
a simply-connected region to a multiply-connected region, and then calculate
the counting function in a multiply-connected region from the heat kernel. The
result of the heat kernel in the multiply-connected region obtained in the
present paper proves Kac's hypothesis. In sections \ref{spectrum} and
\ref{statedensity}, as applications of the counting function, we discuss how
to calculate the asymptotic expressions for spectra and state densities from
$N\left(  \lambda\right)  $. The conclusions are summarized in section
\ref{dc}.

\section{An expression for the counting function: an alternative definition
formula\label{definitionofNL}}

The definition formula for the counting function $N\left(  \lambda\right)  $,
eq. (\ref{NLsgm1}), is not suitable for practical calculations. In fact,
starting from this definition to calculate $N\left(  \lambda\right)  $, one
may encounter many difficulties, e.g., the unsolved problems in number theory
\cite{Berger}. Contrarily, the definition formula for the heat kernel, eq.
(\ref{Kt}), is more operable. In this section, we present an expression for
$N\left(  \lambda\right)  $, which has a similar form with that of the heat
kernel $K\left(  t\right)  $, eq. (\ref{Kt}). Such an expression can be
regarded as an operable definition formula for $N\left(  \lambda\right)  $.

\begin{theorem}%
\begin{equation}
N\left(  \lambda\right)  =\lim_{\beta\rightarrow\infty}\sum_{n}\frac
{1}{e^{\beta\left(  \lambda_{n}-\lambda\right)  }+1}.\label{asymptotic}%
\end{equation}

\end{theorem}

\Proof{
Observing that\[
\lim_{\beta\rightarrow\infty}\frac{1}{e^{\beta\left(  \lambda_{n}-\lambda\right)  }+1}=\left\{
\begin{array}
[c]{cc}1, & \text{ when }\lambda_{n}<\lambda,\\
0, & \text{ when }\lambda_{n}>\lambda,
\end{array}
\right.
\]
we have\[
\lim_{\beta\rightarrow\infty}\sum_{n}\frac{1}{e^{\beta\left(  \lambda
_{n}-\lambda\right)  }+1}=\sum_{\lambda_{n}<\lambda}1=N\left(  \lambda\right)
.
\]
}

Comparing with eq. (\ref{NLsgm1}), eq. (\ref{asymptotic}) is obviously more
operable, since eq. (\ref{asymptotic}) converts the partial sum in eq.
(\ref{NLsgm1}), $\sum_{\lambda_{n}<\lambda}$, into a sum over all possible
values, $\sum_{\lambda_{n}<\infty}$. This will, of course, make the
calculation easier.

\section{The relation between counting function and heat kernel
\label{relation}}

In this section, we provide a relation between the counting function $N\left(
\lambda\right)  $ and the heat kernel $K\left(  t\right)  $. This relation is
an interesting mathematical result and is useful for practical calculations.
As mentioned above, the calculation of the counting function $N\left(
\lambda\right)  $ is more difficult than that of the heat kernel $K\left(
t\right)  $. With the relation between $N\left(  \lambda\right)  $ and
$K\left(  t\right)  $ given in the following, one can first calculate
$K\left(  t\right)  $ and then calculate $N\left(  \lambda\right)  $ from the
result of $K\left(  t\right)  $.

The relation between $N\left(  \lambda\right)  $ and $K\left(  t\right)  $ is
as follows.

\begin{theorem}%
\begin{equation}
K\left(  t\right)  =t\int_{0}^{\infty}N\left(  \lambda\right)  e^{-\lambda
t}d\lambda.\label{K-N}%
\end{equation}

\end{theorem}

\Proof{
The generalized Abel partial summation formula reads\begin{equation}
\sum_{u_{1}<\lambda_{n}\leq u_{2}}b\left(  n\right)  f\left(  \lambda
_{n}\right)  =B\left(  u_{2}\right)  f\left(  u_{2}\right)  -B\left(
u_{1}\right)  f\left(  u_{1}\right)  -\int_{u_{1}}^{u_{2}}B\left(  u\right)
f^{\prime}\left(  u\right)  du, \label{abel}\end{equation}
where $\lambda_{i}\in\mathbb{R}$,
$\lambda_{1}\leq\lambda_{2}\leq\cdots \leq\lambda_{n}\leq\cdots$,
and $\lim_{n\rightarrow\infty}\lambda_{n}=\infty$. $f\left(
u\right)  $ is a continuously differentiable function on $\left[
u_{1},u_{2}\right]  $ $\left(  0\leq u_{1}<u_{2}\text{,
}\lambda_{1}\leq u_{2}\right)  $, $b\left(  n\right)  $ $\left(
n=1,2,3,\cdots\right)  $ are arbitrary complex numbers, and
$B\left(  u\right)  =\sum_{\lambda_{n}\leq u}b\left(  n\right)  $.
We apply the generalized Abel partial summation formula, eq.
(\ref{abel}), with $f\left(  u\right)  =e^{-u\left( s-s_{0}\right)
}$ and $b\left(  n\right)  =a_{n}e^{-\lambda_{n}s_{0}}$, where
$s$, $s_{0}\in\mathbb{C}$. Then\begin{equation}
A\left(  u_{2},s\right)  -A\left(  u_{1},s\right)  =A\left(  u_{2},s_{0}\right)  e^{-u_{2}\left(  s-s_{0}\right)  }-A\left(  u_{1},s_{0}\right)
e^{-u_{1}\left(  s-s_{0}\right)  }+\left(  s-s_{0}\right)  \int_{u_{1}}^{u_{2}}A\left(  u,s_{0}\right)  e^{-u\left(  s-s_{0}\right)  }du,
\label{Au12s}\end{equation}
where\begin{equation}
A\left(  u,s\right)  =\sum_{\lambda_{n}\leq u}a_{n}e^{-\lambda_{n}s}.
\label{Aus}\end{equation}
Setting $a_{n}=1$ in eq. (\ref{Aus}), we find\[
A\left(  \lambda,0\right)  =\sum_{\lambda_{n}\leq\lambda}1=N\left(
\lambda\right)  ,
\]
the counting function, and\[
A\left(  \infty,t\right)  =\sum_{n}e^{-\lambda_{n}t}=K\left(  t\right)  ,
\]
the heat kernel. By eq. (\ref{Aus}), we also have $A\left(
0,t\right)  =0$.
Then, by eq. (\ref{Au12s}), we have\begin{equation}
K\left(  t\right)  =A\left(  \infty,t\right)  -A\left(  0,t\right)  =t\int
_{0}^{\infty}N\left(  \lambda\right)  e^{-\lambda t}d\lambda.
\end{equation}
This is just eq. (\ref{K-N}). }

\begin{theorem}%
\begin{equation}
N\left(  \lambda\right)  =\frac{1}{2\pi i}\int_{c-i\infty}^{c+i\infty}K\left(
t\right)  \frac{e^{\lambda t}}{t}dt,\text{ \ \ }c>\lim_{n\rightarrow\infty
}\frac{\ln n}{\lambda_{n}}.\label{N-K}%
\end{equation}

\end{theorem}

\Proof{
By the Perron formula, we have\begin{equation}
\sum_{\mu_{n}<x}a_{n}=\frac{1}{2\pi i}\int_{c-i\infty}^{c+i\infty}f\left(
t\right)  \frac{x^{t}}{t}dt, \label{Perron}\end{equation}
where\begin{equation}
f\left(  s\right)  =\sum_{n=1}^{\infty}\frac{a_{n}}{\mu_{n}^{s}}, \label{fs}\end{equation}
and $c$ is a constant which is greater than the abscissa of absolute
convergence of the Dirichlet series $f\left(  s\right)  $. Setting
\[
a_{n}=1\text{ and }\mu_{n}=e^{\lambda_{n}}\]
in eq. (\ref{fs}), we obtain the heat kernel,\[
f\left(  t\right)  =\sum_{n=1}^{\infty}e^{-\lambda_{n}t}=K\left(  t\right)  .
\]
The abscissa of absolute convergence of $f\left(  t\right)  $
equals its abscissa of convergence, equaling
$\overline{\lim}_{n\rightarrow\infty}\ln
n/\lambda_{n}=\lim_{n\rightarrow\infty}\ln n/\lambda_{n}$. Thus,
by eq.
(\ref{Perron}), we have\[
N\left(  \lambda\right)  =\sum_{\lambda_{n}<\lambda}1=\frac{1}{2\pi i}\int_{c-i\infty}^{c+i\infty}K\left(  t\right)  \frac{e^{\lambda t}}{t}dt,
\]
and $c>\lim_{n\rightarrow\infty}\ln n/\lambda_{n}$. This proves
the theorem. }

Clearly, the relation between the counting function $N\left(  \lambda\right)
$ and the heat kernel $K\left(  t\right)  $ obtained above is an integral
transformation and the corresponding inverse transformation.

\section{Calculating $N\left(  \lambda\right)  $ from an asymptotics of
$K\left(  t\right)  $\label{calculating}}

In principle, we can calculate the counting function $N\left(  \lambda\right)
$ from the heat kernel $K\left(  t\right)  $ or calculate $K\left(  t\right)
$ from $N\left(  \lambda\right)  $ by the relation (\ref{N-K}) or (\ref{K-N})
directly. However, the relation given above is an integral transformation.
Therefore, when the heat kernel $K\left(  t\right)  $ is in the form of a
series expansion, practically, the above relation is available only when this
integral transformation can be applied to each term of the series, or, in
other words, this series must can be integrated term by term. If the series is
not uniformly convergent, even though the integral of the sum function is
convergent, the integrals of some of the terms may be divergent. In this case,
we need a renormalization procedure to remove the divergences.

\subsection{An asymptotics for $N\left(  \lambda\right)  $}

Concretely, the expansion of heat kernel $K\left(  t\right)  $ can be
expressed as \cite{BEK}%
\begin{equation}
K\left(  t\right)  =\left(  4\pi t\right)  ^{-D/2}\sum_{k=0,\frac{1}%
{2},1,\cdots}^{\infty}B_{k}t^{k},\label{Ktseries}%
\end{equation}
where $B_{k}$ is the heat kernel coefficient, $D$ is the dimension of space.
In fact, in practice only the first several heat kernel coefficients can be
obtained, so we have to integrate term by term. Substituting eq.
(\ref{Ktseries}) into eq. (\ref{N-K}) and exchanging the order of integration
and summation gives%

\begin{equation}
N\left(  \lambda\right)  =\left(  4\pi\right)  ^{-D/2}\sum_{k=0,\frac{1}%
{2},1,\cdots}^{\infty}B_{k}I_{D}\left(  k\right)  ,\label{Nbare}%
\end{equation}
where%
\begin{equation}
I_{D}\left(  k\right)  =\frac{1}{2\pi i}\int_{c-i\infty}^{c+i\infty}%
t^{-D/2+k}\frac{e^{\lambda t}}{t}dt.\label{intDk}%
\end{equation}
Nevertheless, for a power series, unless the series is uniformly convergent,
it is not permissible to integrate term by term, i.e., the order of
integration and summation cannot be exchanged. When calculating an integral of
a series that is not uniformly convergent term by term, one may find that the
integrals of some terms diverge. In our case, it can be directly seen that
only when $k<D/2+1$, $I_{D}\left(  k\right)  $ is convergent. That is to say,
the term with $k\geq D/2+1$ needs to be renormalized for obtaining a finite result.

\subsection{Renormalization}

The integral (\ref{intDk}) is divergent when $k\geq D/2+1$. For obtaining the
counting function $N\left(  \lambda\right)  $ from the series (\ref{Nbare}),
we need to make sense of these divergent integrals. In other words, we need a
renormalization procedure to remove the divergences.

It can be directly seen that eq. (\ref{intDk}) is the inverse Laplace
transformation of the function $t^{-D/2+k-1}$. Thus for $k<D/2+1$, we have
\cite{AS}
\begin{equation}
I_{D}\left(  k\right)  =\frac{1}{2\pi i}\int_{c-i\infty}^{c+i\infty
}t^{-D/2+k-1}e^{\lambda t}dt=\frac{1}{\Gamma\left(  1+D/2-k\right)  }%
\lambda^{D/2-k},\text{ }\left(  k<D/2+1\right)  .\label{intF}%
\end{equation}
The definition domain for the gamma function $\Gamma\left(  1+D/2-k\right)  $
is $k<D/2+1$. Now, we deal with the case of $k\geq D/2+1$. When $-D/2+k-1$ is
a non-negative integer, the integral (\ref{intF}) equals $\delta^{\left(
-D/2+k-1\right)  }\left(  \lambda\right)  $ \cite{Handbook}, where
$\delta^{\left(  n\right)  }\left(  \lambda\right)  $ is the $n$th derivative
of $\delta\left(  \lambda\right)  $. When $-D/2+k-1$ is a positive half
integer, we can redefine the gamma function by analytically continuing the
gamma function via\ the recurrence relation $\Gamma\left(  x\right)  =\frac
{1}{x}\Gamma\left(  x+1\right)  $ \cite{Stone}. Concretely, when defining the
gamma function as $\Gamma\left(  x\right)  =\frac{1}{x}\Gamma\left(
x+1\right)  $, the definition domain of the gamma function is changed to
$k<D/2+2$. Repeating this procedure will eventually analytically continue the
gamma function to the whole real axis.

From the above result we can see that, like that in the perturbative expansion
of the scattering amplitude in quantum field theory, the first several terms
are convergent (like tree diagrams) and the other terms are divergent and
needed to be renormalized (like loop diagrams). For clarity, write the
counting function $N\left(  \lambda\right)  $ in two parts:%
\begin{equation}
N\left(  \lambda\right)  =N_{T}\left(  \lambda\right)  +N_{L}\left(
\lambda\right)  ,
\end{equation}
where%
\begin{equation}
N_{T}\left(  \lambda\right)  =\left(  4\pi\right)  ^{-D/2}\sum_{k=0,\frac
{1}{2},1,\cdots}^{\frac{D}{2}+\frac{1}{2}}B_{k}I_{D}\left(  k\right)
\label{NT}%
\end{equation}
is the convergent part and%
\begin{equation}
N_{L}\left(  \lambda\right)  =\left(  4\pi\right)  ^{-D/2}\sum_{k=\frac{D}%
{2}+1,\frac{D}{2}+\frac{3}{2},\cdots}^{\infty}B_{k}I_{D}^{R}\left(  k\right)
\label{NL}%
\end{equation}
is the renormalized divergent part, where $I_{D}^{R}\left(  k\right)  $
denotes the renormalized result. $N_{L}\left(  \lambda\right)  $ is only a
higher-order contribution to the counting function and is often negligible.

After renormalization, eqs. (\ref{NT}) and (\ref{NL}) can be written as%
\begin{align}
N\left(  \lambda\right)   & =\sum_{k=0,\frac{1}{2},1,\cdots}^{\infty}%
C_{k}\lambda^{D/2-k}+\sum_{l=0,1,2,\cdots}^{\infty}\left(  4\pi\right)
^{-D/2}B_{1+D/2+l}\delta^{\left(  l\right)  }\left(  \lambda\right)
\nonumber\\
& =\sum_{k=0,\frac{1}{2},1,\cdots}^{\frac{D}{2}}C_{k}\lambda^{D/2-k}%
+\sum_{k=\frac{D}{2}+\frac{1}{2},\frac{D}{2}+\frac{3}{2},\cdots}^{\infty}%
C_{k}\lambda^{D/2-k}+\sum_{k=\frac{D}{2}+1,\frac{D}{2}+2,\cdots}^{\infty
}\left(  4\pi\right)  ^{-D/2}B_{k}\delta^{\left(  k-\left(  \frac{D}%
{2}+1\right)  \right)  }\left(  \lambda\right) \label{NLbyCk}%
\end{align}
with the counting function coefficient%
\begin{equation}
C_{k}=\left(  4\pi\right)  ^{-D/2}\frac{B_{k}}{\Gamma\left(  1+D/2-k\right)
}.\label{Ck}%
\end{equation}
It should be emphasized that the delta-function terms in eq. (\ref{NLbyCk})
will contribute the zero-point energy; this is, such contributions will play
an essentially important role in the calculation of the Casimir effect.

\section{Comparison of the two approaches for calculating $N\left(
\lambda\right)  $: $D$-dimensional boxes \label{D-box}}

The results presented in sections \ref{definitionofNL} and \ref{relation} show
that there are two approaches for calculating the counting function $N\left(
\lambda\right)  $. The first is to directly calculate $N\left(  \lambda
\right)  $ by eq. (\ref{asymptotic}), the operable expression of $N\left(
\lambda\right)  $ presented in this paper; the second is to first calculate
the heat kernel $K\left(  t\right)  $ and then to calculate $N\left(
\lambda\right)  $ from $K\left(  t\right)  $ by the relation between $N\left(
\lambda\right)  $ and $K\left(  t\right)  $. In this section, as an example,
we calculate $N\left(  \lambda\right)  $ by the above two approaches
respectively for a rectangular $D$-dimensional box.

The spectrum of the minus Laplace operator of a $D$-dimensional rectangle box reads%

\begin{equation}
\lambda_{n_{1},n_{2},\cdots,n_{D}}=\pi^{2}\left(  \frac{n_{1}^{2}}{L_{1}^{2}%
}+\frac{n_{2}^{2}}{L_{2}^{2}}+\cdots+\frac{n_{D}^{2}}{L_{D}^{2}}\right)
,~~n_{i}=1,2\cdots,
\end{equation}
where $L_{i}$ is the side length.

First, we directly calculate the counting function $N\left(  \lambda\right)  $
from eq. (\ref{asymptotic}):%
\begin{equation}
N\left(  \lambda\right)  =\lim_{\beta\rightarrow\infty}\sum_{n_{1}%
,n_{2},\cdots,n_{D}}\frac{1}{e^{\beta\left(  \lambda_{n_{1},n_{2},\cdots
,n_{D}}-\lambda\right)  }+1}.
\end{equation}
By the Euler-MacLaurin formula $\sum\limits_{n=0}^{\infty}F(n)=\int
_{0}^{\infty}F(n)dn+\frac{1}{2}F(0)-\frac{1}{2!}B_{2}F^{\prime}(0)-\frac
{1}{4!}B_{4}F^{\prime\prime\prime}(0)+\cdots$, where $B_{2}$, $B_{4}$,
$\cdots$ are Bernoulli numbers, we achieve%
\begin{align}
N\left(  \lambda\right)   &  =\left(  4\pi\right)  ^{-D/2}\left[  \frac
{V}{\Gamma\left(  D/2+1\right)  }\lambda^{D/2}-\sum_{i=1}^{D}\frac{V}{L_{i}%
}\frac{\sqrt{\pi}}{\Gamma\left(  D/2+1/2\right)  }\lambda^{D/2-1/2}\right.
\nonumber\\
&  +\left.  \sum_{i<j}^{D}\frac{V}{L_{i}L_{j}}\frac{\pi}{\Gamma\left(
D/2\right)  }\lambda^{D/2-1}+\cdots+\left(  -1\right)  ^{D}\pi^{D/2}\right]
,\label{NfromK}%
\end{align}
where $V=$ $L_{1}L_{2}\cdot\cdots\cdot L_{D}$ is the volume.

Alternatively, we can first calculate the heat kernel from the definition
formula of $K\left(  t\right)  $,
\begin{equation}
K\left(  t\right)  =\sum_{n_{1},n_{2},\cdots,n_{D}}e^{-\lambda\left(
n_{1},n_{2},\cdots,n_{D}\right)  t}.
\end{equation}
A direct calculation by the Euler-MacLaurin formula gives the heat kernel
coefficients:%
\begin{align}
B_{0}  &  =V,\ B_{1/2}=-\sqrt{\pi}\sum_{i=1}^{D}\frac{V}{L_{i}},\ B_{1}%
=\pi\sum_{i<j}^{D}\frac{V}{L_{i}L_{j}},\cdots,\nonumber\\
B_{\nu/2}  &  =\left(  -1\right)  ^{\nu}\pi^{\nu/2}\sum_{i_{1}<i_{2}%
<\cdots<i_{\nu}}^{D}\frac{V}{L_{i_{1}}L_{i_{2}}\cdots L_{i_{\nu}}%
}.\label{Bkbox}%
\end{align}
By the relation between the counting function coefficient and the heat kernel
coefficient, eq. (\ref{Ck}), we also achieve eq. (\ref{NfromK}).

From the above result, we can see the equivalence between these two approaches.

\section{Calculating $N\left(  \lambda\right)  $ from $K\left(  t\right)  $:
three-dimensional balls \label{3Dballs}}

In this section, as an example, we calculate the counting function $N\left(
\lambda\right)  $ from the series expansion of the heat kernel $K\left(
t\right)  $ for a three-dimensional ball. It should be emphasized that for
calculating the heat kernel $K\left(  t\right)  $, it is not needed to know
the eigenvalue spectrum in advance \cite{BEK}.

For a three-dimensional ball, from eq.(\ref{Ck}), the expansion coefficients
of the counting function $N\left(  \lambda\right)  $ can be obtained as%
\begin{equation}
C_{k}=\left(  4\pi\right)  ^{-3/2}\frac{B_{k}}{\Gamma\left(  5/2-k\right)
}.\label{Ck3ball}%
\end{equation}
The heat kernel coefficients for Dirichlet, Neumann, and Robin boundary
conditions for three-dimensional balls are calculated explicitly in \cite{BEK}.

Take the case of Dirichlet boundary condition as an example. Ref. \cite{BEK}
gives the first $21$ heat kernel coefficients. From this, we can obtain the
first $21$ counting function coefficients $C_{k}$ by eq. (\ref{Ck3ball}). Only
taking the tree-diagram-like part into account, from eq. (\ref{NT}), we have
\begin{equation}
N\left(  \lambda\right)  =\frac{2}{9\pi}R^{3}\lambda^{3/2}-\frac{1}{4}%
R^{2}\lambda+\frac{2}{3\pi}R\lambda^{1/2}-\frac{1}{48}-\frac{2}{315\pi}%
\frac{1}{R\lambda^{1/2}}.\label{NT3Ball}%
\end{equation}
It can be directly checked that the tree-diagram-like part provides the main
contribution to the $N\left(  \lambda\right)  $.

\section{Multiply-connected cases: two-dimensional $K\left(  t\right)  $ and
$N\left(  \lambda\right)  $\label{2Dmul}}

An interesting special case is about two-dimensional counting function and
two-dimensional heat kernel, which is just the original problem formulated by
Kac as "Can one hear the shape of a drum?". The first result of this problem
is proved by Weyl \cite{Weyl} and then improved by Pleijel \cite{Pleijel} and
Kac \cite{Kac}. In \cite{Kac}, Kac pointed out a special relation between
$N\left(  \lambda\right)  $ and $K\left(  t\right)  $: in two dimensions, the
leading-order term of $N\left(  \lambda\right)  $ is accidentally equal to the
leading-order term of $K\left(  t\right)  $. However, beyond the leading
order, there is no further result. Moreover, for the multiply-connected case,
Kac made a hypothesis about the topological contribution on the heat kernel
$K\left(  t\right)  $: the contribution from the nontrivial topology
(connectivity) is in proportion to the Euler-Poincar\'{e} characteristic number.

In this section, (1) with the help of the relation between $N\left(
\lambda\right)  $ and $K\left(  t\right)  $ given above, we give a proof for
Kac's result: the leading-order term of $N\left(  \lambda\right)  $ equals the
leading-order term of $K\left(  t\right)  $; (2) based on the Gauss-Bonnet
theorem, we first generalize the simply-connected heat kernel $K\left(
t\right)  $ given in \cite{NPD} to the multiply-connected case (this result
proves Kac's conjecture: in two dimensions, the topology contribution is
proportion to the Euler-Poincar\'{e} characteristic number), and, then,
calculate the two-dimensional\ $N\left(  \lambda\right)  $ in a
multiply-connected region with the help of the relation between $N\left(
\lambda\right)  $ and $K\left(  t\right)  $.

\subsection{The leading contribution of $N\left(  \lambda\right)  $ and
$K\left(  t\right)  $ in two dimensions}

Using the relation between $N\left(  \lambda\right)  $ and $K\left(  t\right)
$, we first prove that when the number of eigenstates per unit interval (the
density of eigenstates) $\rho\left(  \lambda\right)  $ is a constant,
$N\left(  \lambda\right)  $ equals $K\left(  t\right)  $ in the limit
$\lambda\rightarrow\infty$ or $t\rightarrow0$, i.e., in the limit
$\lambda\rightarrow\infty$ or $t\rightarrow0$,%
\begin{equation}
N\left(  \lambda\right)  =K\left(  \frac{1}{\lambda}\right)  \text{\ or
\ }N\left(  \frac{1}{t}\right)  =K\left(  t\right)  .\label{lttendsinf}%
\end{equation}

The proof is straightforward. In the limit $\lambda\rightarrow\infty$ or
$t\rightarrow0$, the summations can be converted into integrals:%
\begin{align}
N\left(  \lambda\right)   &  =\sum_{\lambda_{n}<\lambda}1=\int_{0}^{\lambda
}\rho\left(  \lambda^{\prime}\right)  d\lambda^{\prime},\\
K\left(  t\right)   &  =\sum_{n}e^{-\lambda_{n}t}=\int_{0}^{\infty}\rho\left(
\lambda^{\prime}\right)  e^{-\lambda^{\prime}t}d\lambda^{\prime}.
\end{align}
If $\rho\left(  \lambda\right)  =C$, where $C$ is a constant, then%
\begin{align}
N\left(  \lambda\right)   &  =C\lambda,\\
K\left(  t\right)   &  =\frac{C}{t}.
\end{align}
This proves eq. (\ref{lttendsinf}).

In two dimensions, the state density for the eigenstate of the minus Laplace
operator is a constant, so the leading contribution of the counting function
equals the leading contribution of the heat kernel. This is just the case
appeared in Kac's work \cite{Kac}.

\subsection{$K\left(  t\right)  $ and $N\left(  \lambda\right)  $ in
Multiply-connected regions}

For achieving the counting function in a multiply-connected region by the
relation between $K\left(  t\right)  $ and $N\left(  \lambda\right)  $, we
need to start with the heat kernel in a multiply-connected region. For
two-dimensional heat kernels, there are two known results: in the
simply-connected case, ref. \cite{NPD} provides a series expansion of the heat
kernel; in the multiply-connected case, Kac makes a hypothesis on the heat
kernel \cite{Kac}. In the following, based on the Gauss-Bonnet theorem, we
first generalize the simply-connected result of heat kernel $K\left(
t\right)  $ given in \cite{NPD} to the multiply-connected case. The result can
be viewed as a proof of Kac's hypothesis.

In the simply-connected case, the heat kernel for a two-dimensional plane
bounded by a smooth curve $\Gamma_{s}$ and with the Dirichlet boundary
condition on $\Gamma_{s}$ reads \cite{NPD}%
\begin{equation}
K\left(  t\right)  =\frac{S}{4\pi t}-\frac{L}{8\sqrt{\pi t}}+\frac{1}{12\pi
}\int_{\Gamma_{s}}k\left(  s\right)  ds+\cdots,\label{Kt2Dsimply}%
\end{equation}
where $S$ is the area of the region, $L$ is the length of $\Gamma_{s}$,
$k\left(  s\right)  $ is the curvature of the curve $\Gamma_{s}$ at the point
$s $.

\begin{figure}[ptb]
\begin{center}
\includegraphics[width=5in]{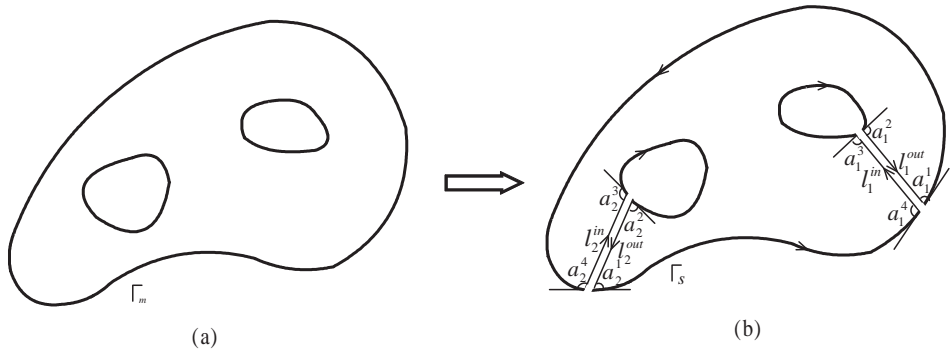} \label{fig}
\end{center}
\caption{Converting a multiply-connected region to a simply-connected region.
}%
\end{figure}

To generalize the result of the simply-connected region, eq. (\ref{Kt2Dsimply}%
), to the multiply-connected case, we first convert the multiply-connected
region bounded by $\Gamma_{m}$ (figure 1a) to a simply-connected one bounded
by a piecewise smooth simple closed curve $\Gamma_{s}$ (figure 1b). In the
simply-connected region illustrated in figure 1b, the first two terms which
are proportional to the area and the perimeter of the region, respectively,
can be calculated by the method in the simply-connected case directly. For the
third term, by the Gauss-Bonnet theorem, we can calculate the integral of the
curvature along $\Gamma_{s}$. The Gauss-Bonnet theorem reads \cite{Chern}
\begin{equation}%
{\displaystyle\sum\limits_{i}}
\left(  \pi-a_{i}\right)  +\int_{\Gamma_{s}}k\left(  s\right)  ds+%
{\displaystyle\iint}
Kd\sigma=2\pi\chi,\label{GB}%
\end{equation}
where $a_{i}$ is the interior angle of $\Gamma_{s}$ at each vertex, $K=0$ in
the present case is the Gauss curvature, and $\chi$ is the Euler-Poincar\'{e}
characteristic number. From figure 1, since the integrals along $l_{i}^{in}$
and $l_{i}^{out}$ ($i=1,\cdots,r$) cancel each other in pairs, we achieve%
\begin{equation}
\int_{\Gamma_{m}}k\left(  s\right)  ds=\int_{\Gamma_{s}}k\left(  s\right)
ds,\label{mtos}%
\end{equation}
where $r$ is the number of holes in the region. Moreover, it is also easy to
see that%
\begin{equation}%
{\displaystyle\sum\limits_{i=1}^{r}}
{\displaystyle\sum\limits_{\alpha=1}^{4}}
\left(  \pi-a_{i}^{\alpha}\right)  =r2\pi.
\end{equation}
The region bounded by $\Gamma_{s}$ in figure 1b is simply-connected, so
$\chi=1$. Thus, from eq. (\ref{GB}), we have%
\begin{equation}
\int_{\Gamma_{s}}k\left(  s\right)  ds=2\pi\left(  1-r\right)  .\label{intr}%
\end{equation}
By the approach given by \cite{MV}, we can calculate the first two terms.
Then, with eqs. (\ref{Kt2Dsimply}), (\ref{mtos}), and (\ref{intr}), we have%
\begin{equation}
K\left(  t\right)  =\frac{S}{4\pi t}-\frac{L}{8\sqrt{\pi t}}+\frac{1-r}%
{6}.\label{Ktmul}%
\end{equation}
This is just the result that Kac hypothesized in \cite{Kac}.

By the relation between $N\left(  \lambda\right)  $ and $K\left(  t\right)  $,
we can directly calculate the counting function in the two-dimensional
multiply-connected region:%
\begin{equation}
N\left(  \lambda\right)  =\frac{S}{4\pi}\lambda-\frac{L}{\pi}\sqrt{\lambda
}+\frac{1-r}{6}.
\end{equation}

\section{Calculating spectrum from $N\left(  \lambda\right)  $\label{spectrum}%
}

From the counting function $N\left(  \lambda\right)  $, we can directly obtain
the asymptotic expression for the spectrum of the system \cite{CH}. In view of
the fact that one can obtain the heat kernel $K\left(  t\right)  $ and,
accordingly, obtain the counting function $N\left(  \lambda\right)  $ without
knowing the spectrum in advance \cite{BEK}, this result can serve as an
approach for calculating the spectrum.

The counting function is the number of eigenvalues smaller than $\lambda$, so
for the $n$-th eigenvalue $\lambda_{n}$, we have%
\begin{equation}
N\left(  \lambda_{n}\right)  =n.\label{NLbdeqn}%
\end{equation}
Therefore, from eqs. (\ref{NLbdeqn}) and (\ref{NLbyCk}), we have%
\begin{equation}
\left(  4\pi\right)  ^{-D/2}\sum_{k=0,\frac{1}{2},1,\cdots}^{\infty}%
\frac{B_{k}}{\Gamma\left(  1+D/2-k\right)  }\lambda_{n}^{D/2-k}%
=n.\label{NLbdR}%
\end{equation}
The eigenvalue $\lambda_{n}$ can be solved from eq. (\ref{NLbdR}).

For approximately solving eq. (\ref{NLbdR}), we assume that $\lambda_{n}$ can
be expanded as%
\begin{equation}
\lambda_{n}=%
{\displaystyle\sum\limits_{m=0}^{\infty}}
\alpha_{m}n^{\left(  2-m\right)  /D}.\label{Lbdn}%
\end{equation}
Then we have%
\begin{align}
\lambda_{n}  &  =4\pi\Gamma^{2/D}\left(  \frac{D}{2}+1\right)  \frac{1}%
{B_{0}^{2/D}}n^{2/D}-4\sqrt{\pi}\frac{\Gamma^{1+1/D}\left(  \frac{D}%
{2}+1\right)  }{D\Gamma\left(  \frac{D}{2}+\frac{1}{2}\right)  }\frac{B_{1/2}%
}{B_{0}^{1+1/D}}n^{1/D}\nonumber\\
&  +\left[  \frac{D\Gamma^{2}\left(  \frac{D}{2}\right)  }{4\Gamma^{2}\left(
\frac{D}{2}+\frac{1}{2}\right)  }\frac{B_{1/2}^{2}}{B_{0}^{2}}-\frac{B_{1}%
}{B_{0}}\right]  n^{0}+\cdots.
\end{align}

As examples, we list some spectra obtained by this approach for various
dimensional balls with Dirichlet, Neumann, and Robin boundary conditions,
respectively. The heat kernel coefficients are given in \cite{BEK}.

For a three-dimensional ball, the spectrum for the Dirichlet and the Neumann
or Robin boundary conditions is%

\begin{equation}
\lambda_{n}=\frac{3}{2}\left(  6\pi^{2}\right)  ^{1/3}\frac{1}{R^{2}}%
n^{2/3}\pm\frac{3}{8}\left(  6\pi^{2}\right)  ^{2/3}\frac{1}{R^{2}}%
n^{1/3}+\left(  \frac{27}{64}\pi^{2}-2\right)  \frac{1}{R^{2}},
\end{equation}
where $R$ is the radius. In this equation and following, the upper sign stands
for the Dirichlet boundary condition and the lower sign for the Neumann or
Robin boundary condition.

For a four-dimensional ball, the spectrum is%

\begin{equation}
\lambda_{n}=\frac{8}{R^{2}}n^{1/2}\pm\frac{16\sqrt{2}}{3}\frac{1}{R^{2}%
}n^{1/4}-\frac{26}{9}\frac{1}{R^{2}},
\end{equation}
and for a five-dimensional ball, the spectrum is%

\begin{equation}
\lambda_{n}=\frac{1}{2}\left(  450\sqrt{2}\pi\right)  ^{2/5}\frac{1}{R^{2}%
}n^{2/5}\pm\frac{15\pi}{32}\left(  3600\pi\right)  ^{1/5}\frac{1}{R^{2}%
}n^{1/5}+\left(  \frac{1125}{1024}\pi^{2}-\frac{20}{3}\right)  \frac{1}{R^{2}%
}.
\end{equation}

\section{The state density\label{statedensity}}

In this section, we provide two approaches to achieve the state density based
on the above results.

\subsection{Calculating the state density from $N\left(  \lambda\right)  $ and
$K\left(  t\right)  $}

The first approach is straightforward. The meaning of the counting function
$N\left(  \lambda\right)  $ is the number of the states whose eigenvalues are
smaller than $\lambda$, so the state density reads%
\begin{equation}
\rho\left(  \lambda\right)  =\frac{dN\left(  \lambda\right)  }{d\lambda
}.\label{density}%
\end{equation}
Thus, from eq. (\ref{NLbyCk}), we have%
\begin{align}
\rho\left(  \lambda\right)   & =\left(  4\pi\right)  ^{-D/2}\sum_{k=0,\frac
{1}{2},1,\cdots}^{\infty}\frac{B_{k}}{\Gamma\left(  D/2-k\right)  }%
\lambda^{D/2-k-1}+\sum_{l=0,1,2,\cdots}^{\infty}\left(  4\pi\right)
^{-D/2}B_{1+D/2+l}\delta^{\left(  l+1\right)  }\left(  \lambda\right)
.\nonumber\\
& =\left(  4\pi\right)  ^{-D/2}\sum_{k=0,\frac{1}{2},1,\cdots}^{\frac{D}%
{2}-\frac{1}{2}}\frac{B_{k}}{\Gamma\left(  D/2-k\right)  }\lambda
^{D/2-k-1}+\left(  4\pi\right)  ^{-D/2}\sum_{k=\frac{D}{2}+\frac{1}{2}%
,\frac{D}{2}+\frac{3}{2},\cdots}^{\infty}\frac{B_{k}}{\Gamma\left(
D/2-k\right)  }\lambda^{D/2-k-1}\nonumber\\
& +\sum_{k=\frac{D}{2}+1,\frac{D}{2}+2,\cdots}^{\infty}\left(  4\pi\right)
^{-D/2}B_{k}\delta^{\left(  k-\frac{D}{2}\right)  }\left(  \lambda\right)
\label{sd}%
\end{align}
From this result, one can obtain the state density once he knows the heat
kernel coefficients.

\subsection{Calculating the state density from the spectrum}

An alternative way for obtaining the state density is to start with the
expression of the counting function given in section \ref{definitionofNL}.

Derivating both sides of eq. (\ref{asymptotic}) by $\lambda$ gives%
\begin{equation}
\rho\left(  \lambda\right)  =\frac{dN\left(  \lambda\right)  }{d\lambda}%
=\lim_{\beta\rightarrow\infty}\sum_{n}\frac{\beta e^{\beta\left(  \lambda
_{n}-\lambda\right)  }}{\left[  e^{\beta\left(  \lambda_{n}-\lambda\right)
}+1\right]  ^{2}}.\label{density2}%
\end{equation}
Or, approximately, we can convert the summation over $n$ to an integral:%
\begin{equation}
\rho\left(  \lambda\right)  =\lim_{\beta\rightarrow\infty}%
{\displaystyle\int}
dn\frac{\beta e^{\beta\left(  \lambda_{n}-\lambda\right)  }}{\left[
e^{\beta\left(  \lambda_{n}-\lambda\right)  }+1\right]  ^{2}}.
\end{equation}
Then, we can directly calculate the state density from the spectrum.

\section{Conclusions\label{dc}}

In this paper, we reveal a relation between the counting function $N\left(
\lambda\right)  $ and the heat kernel $K\left(  t\right)  $ and provide an
operable expression for $N\left(  \lambda\right)  $. By the relation, one can
calculate the counting function from a known heat kernel, and \textit{vice
versa}. By the expression of the counting function presented in this paper,
one can achieve the counting function by a direct calculation.

The relation between $N\left(  \lambda\right)  $ and $K\left(  t\right)  $ is
an integral transformation and its inverse transformation. When calculating
$N\left(  \lambda\right)  $ from $K\left(  t\right)  $ by this relation,
however, one may encounter the problem of divergence. This is because in most
cases the expression of $K\left(  t\right)  $ is in the form of a power series
(in fact, often only first several heat kernel coefficients can be obtained),
but the series is not uniformly convergent. As a result, when applying the
integral transformation to the series, one needs to integrate term by term.
However, to integrate term by term is not feasible: the integration of some
terms will diverge. For removing the divergences, we develop a renormalization procedure.

The results of this paper provide two approaches for calculating the counting
functions: one is to calculate $N\left(  \lambda\right)  $ directly from the
expression of $N\left(  \lambda\right)  $ given in section
\ref{definitionofNL} and the other is based on the relation between $N\left(
\lambda\right)  $ and $K\left(  t\right)  $ given in section \ref{relation}.
As a comparison between the two approaches, we calculate the counting function
for $D$-dimensional boxes by these two approaches, respectively.

As applications of the relation between $N\left(  \lambda\right)  $ and
$K\left(  t\right)  $, we also calculate the counting functions for
three-dimensional balls and two-dimensional multiply-connected irregular
regions. For calculating the counting function of a two-dimensional
multiply-connected region, we generalize the result of the heat kernel of a
two-dimensional simply-connected region to a multiply-connected one, and the
result proves Kac's conjecture on the two-dimensional multiply-connected heat kernel.

Moreover, we also present approaches for calculating eigenvalue spectra and
state densities from the counting function.

\acknowledgments{We are very indebted to Dr. G. Zeitrauman for his
encouragement. This work is supported in part by NSF of China under
Grant No. 10605013 and the Hi-Tech Research and Development
Programme of China under Grant No. 2006AA03Z407.}

\end{document}